**Title:**

**Migration-selection balance at multiple loci and selection on dominance and recombination**

Alexey Yanchukov

Ecology, Evolution and Marine Biology, University of California Santa Barbara

Santa Barbara, CA 93106-9620

USA

Stephen R. Proulx

Ecology, Evolution and Marine Biology, University of California Santa Barbara

Santa Barbara, CA 93106-9620

USA

Running title: Evolutionary consequences of gene introgression

Corresponding author:

Alexey Yanchukov

Email: yawa33@gmail.com

Phone: +1 805 893 4086

Ecology, Evolution and Marine Biology, University of California Santa Barbara

Life Scinces Building, room 3203

Santa Barbara, CA 93106-9620

USA

ABSTRACT

A steady influx of a single deleterious multilocus genotype will impose genetic load on the resident population and leave multiple descendants carrying various numbers of the foreign alleles. Provided that the foreign types are rare at equilibrium, and all immigrant genes are eventually eliminated by selection, the population structure can be inferred explicitly from the branching process taking place within a single immigrant lineage. Unless the migration and recombination rates were high, this simple method was a close approximation to the simulation with all possible multilocus genotypes considered. Once the load and the foreign genotypes frequencies are known, it becomes possible to estimate selection acting on the invading modifiers of (i) dominance and (ii) recombination rate on the foreign gene block. We found that the modifiers of the (i) type are able to invade faster than the type (ii) modifier, however, this result only applies in the strong selection / low migration / low recombination scenario. Varying the number of genes in the immigrant genotype can have a non-monotonic effect on the migration load and the modifier's invasion rate: although blocks carrying more genes can give rise to longer lineages, they also experience stronger selection pressure. The heaviest load is therefore imposed by the genotypes carrying moderate numbers of genes.

INTRODUCTION

Gene flow is a major force shaping the evolution of closely related sexual populations, which can prevent local adaptation (BLANQUART *et al.* 2012; STORFER *et al.* 1999) or facilitate generalism (GRAY and GODDARD 2012; SEEHAUSEN 2004), erase or maintain intraspecific polymorphism (RONCE AND KIRKPATRICK 2001; STAR ET AL. 2007), promote the evolution of female mating preferences (PROULX 2001; VAN DOORN AND WEISSING 2006) and lead to speciation through reinforcement (SERVEDIO AND KIRKPATRICK 1997; SERVEDIO AND NOOR 2003). Such diverse evolutionary outcomes arise because the exchange of genetic material generates variance and so is intervened by some form of selection: in a vast number of studied cases, selection affects only a discrete set of genes, small in proportion to the individual genome size (NADEAU *et al.* 2012; RIESEBERG *et al.* 1999; TURNER *et al.* 2005). The dynamics of gene flow and selection in such multilocus system are well illustrated in a one-way migration model: at the point of first entry into the resident population, the set of foreign genes is intact (i.e. is in complete linkage disequilibrium, LD), but after repeated backcrosses with the resident genotypes (i.e. introgression), there will be multiple descendants of the first migrants carrying some proportion of the initial foreign genotype. Because natural selection acts on the individual level, the actual selective pressure on each foreign gene will depend on its association with the other selected loci: in the most common scenario where the foreign genes are deleterious, the first generation hybrids with high LD will be strongly selected against, while the backcrosses will experience less selective pressure (HARRISON 1993).

The interplay between gene flow and selection thus leaves a specific signature on the genetic structure of the resident population, observed, for example, in many hybrid zones (JIGGINS AND MALLET 2000) and subject to extensive applied modeling aimed primarily at the data analysis (SZYMURA AND BARTON 1991). More rigorous theoretical treatment of the introgression, is, however, difficult since too many genotypes have to be considered even for a moderate number of loci: significant progress in this area has been made but involves a great deal of approximation and simplifying assumptions about the nature of recombination of the foreign set of genes (BARTON 1983; BARTON AND BENGTSSON 1986). Here, we develop this theory further and propose a new simple model of gene flow and selection that allows for an explicit characterization of the resident population structure, provided that the population is large, migration rate is low and the foreign genes are evenly distributed in a linear genome block.

Under the premise that most non-neutral variance between populations is generated by diversifying selection, one expects that genes brought into the new environment (including new genomic environment) will experience negative selection pressure (BOLNICK AND NOSIL 2007). This means that all *selected* introgressed material descending from the first immigrant individual will almost certainly disappear as time progresses, but until then, the mean fitness of the resident population will remain suboptimal. We show that it is possible to calculate the reduction in population mean fitness caused by the immigrant lineage throughout its path to extinction. In fact, this measure turns out to be equal to the actual migration load imposed on the population by gene flow and calculated at any time point once the migration-selection process is at equilibrium, because immigrant lineages, arriving at a slow rate, will segregate *independently* in a large resident population. Based on the same principle, we derive the equilibrium frequencies of genotype classes carrying the same numbers of the introgressing genes, and thus arrive at a good approximation of the population genetic structure. Once the population structure corresponding to a given point in parameter space is known, we can estimate the strength of selection favoring specific microevolutionary mechanisms that render the migration-selection balance unstable (OTTO AND DAY 2011).

What theoretical consequences can follow from the equilibrium between multilocus gene flow and selection? Since mean population fitness is reduced, any novel mechanism that allows it to recover towards the maximal level should be favored by selection (PROULX AND PHILLIPS 2005; PROULX AND SERVEDIO 2009). Obviously and trivially, a reduction in the migration load would be achieved simply by decreasing gene flow, for example by reducing the number of migrants between demes (BILLIARD AND LENORMAND 2005). However, this could also be achieved by selection on mating preferences (PROULX 2001; PROULX and SERVEDIO 2009; SERVEDIO 2007). Here we add to the vast body of literature on this subject by asking what level of selection pressure will act on a modifier that (i) masks the deleterious effect of the immigrant genes, therefore increasing the robustness of the resident genotype to the gene flow and (ii) suppresses recombination between the foreign genes, thus increasing the efficacy of selection on multiple loci. Canalization of deleterious alleles is widespread among eukaryotes, particularly in the form of diploid dominance, and a few theoretical studies demonstrated that either of these can evolve in response to migration (OTTO AND BOURGUET 1999; YANCHUKOV AND PROULX 2012). Genomic clustering of the adaptive loci has also been shown to evolve under the migration-selection balance (BANK *et al.* 2012; YEAMAN and WHITLOCK 2011). In this article, we demonstrate that the corresponding dominance or recombination modifiers can increase in frequency under comparable conditions, and that the amount of selection acting on them can be significant relative to the migration load (PROULX AND PHILLIPS 2005).

THE GENERAL MODEL: MULTILOCUS INTROGRESSION AT THE MIGRATION-SELECTION EQUILIBRIUM.

Our basic model is a generalization of Barton's (1983) model of low rate gene exchange between two demes. Consider a large population in which a small proportion *m* of residents is replaced by an equal fraction of immigrants each generation. All immigrants have the same genotype consisting of a finite number of discrete genes, that is, the genomic distance between any two genes far exceeds the size of the gene itself, and recombination is unlikely within the gene. Following migration, there is direct selection against the foreign genotypes in the resident population, which generally becomes weaker as the average number of the foreign genes per individual decreases due to recombination, but remains strong enough to counter-balance gene flow and maintain a migration-selection equilibrium. The total fraction of recombinant individuals is still assumed to be small even after the equilibrium is reached, so that we can safely ignore unions between gametes that both carry foreign genes. Selection is therefore essentially haploid: even if the immigrants were diploid, they would appear as haploid gametes after just one round of selection. Let $\Pr(i \rightarrow j)$ be the unspecified (for now) probability of transition between the non-identical genotypes labeled $i$ and $j$, carrying the numbers $k$ and $l$ of foreign genes, respectively. Note that because immigrants only mate with the residents, $j$ cannot carry more foreign genes than $i$, and let $\pi$ be the initial immigrant type, which has the maximum number of genes $k_\pi$ (Fig. 1). Over time, this transition probability will completely determine the distribution of genotypes within a set of individuals sharing a common ancestry and carrying the introgressed genetic material, which we will call a *lineage* (BAIRD ET AL. 2003). We will now demonstrate that knowledge of the mechanisms of selection and recombination that takes place within a single lineage is sufficient to characterize the genetic structure of the population at migration-selection equilibrium.

**Mean population fitness:** Let us examine a lineage $A_\pi$ descending from a single initial immigrant genotype $\pi$ introduced at the generation $t_0$, and assume for simplicity that exactly one immigrant arrives every generation. A subset $A_\pi(t) \subseteq A_\pi$ will represent all members of the lineage present after migration and before selection, at each of the successive generations following the introduction: $t = 1,2,3..t_{ext}$, where $t_{ext}$ is the generation when the last member of the lineage becomes deterministically extinct. The ultimate extinction of the lineage is guaranteed by our previous assumption that recombination cannot break up the foreign genotypes beyond a single gene; once this limit has been reached; the single gene is guaranteed to eventually be eliminated by selection (AGRAWAL and WHITLOCK 2012; HALDANE 1937; HALDANE 1957). Because the number of generations after which the lineage is sampled, $t$, is counted relative to the time of introduction $t_0$, and the migration-selection processes is at equilibrium, varying the time of introduction will have the same effect as varying the time of sampling. It is easy to see that in our model, where no interaction between the foreign genotypes is assumed, population at any time can be represented as a set of *independent* lineages with different times of origin, but otherwise identical (Fig. 2). Moreover, since the initial immigrant genotype is introduced at the same rate every generation, the set $A(t) \supseteq A_\pi(t)$ of all lineages sampled at time $t$ will be equivalent to the set $A_\pi$ representing a single lineage:

$$A(t) = A_\pi \qquad (1)$$

Eq (1) will hold for any number of immigrants arriving each generation, as long as the migration rate $m$ is constant. It is thus possible to characterize the population by a branching process (BP) that takes place within just a single lineage (BAIRD 1995; BAIRD *et al.* 2003; HEATHCOTE 1965). The left-hand side of (1) can be re-written in the frequency representation, as a sum $\sum_i p_i w_i$, where $p_i$ is the frequency of the genotype $i$ in the current generation after migration but before selection and $w_i$ is genotype fitness. Adding the fraction of the resident genotype after migration, $1-m$, we obtain the expression for the mean population fitness, $\bar{w}$:

$$\bar{w} = 1 - m + \sum_i p_i w_i \qquad (2)$$

We must now re-write the right-hand side of (1) in terms of the relative frequencies of all the genotypes within a lineage, sampled at different time points. It will include the frequency of the initial genotype $\pi$ that immigrated and have been selected against in the current generation, $w_\pi m$, minus the fraction of the population mean fitness, $\bar{w}$, occupied by the progeny of $\pi$ destined to become extinct in the future (Fig. 3A):

$$\bar{w} = 1 - m + w_\pi m \left( 1 - \frac{\sum_j \Pr(\pi \to j) Z_j + \Pr(\pi \to \pi) Z_\pi}{\bar{w}} \right) \qquad (3)$$

The term $Z_j$, which we will call the relative *size* of the lineage $j$, represents the contribution to the mean fitness made by all future descendants of the genotype $j$. Correspondingly, $Z_\pi$ is the size of the lineage descending from $\pi$ that survived the first round of selection and did not recombine. The dynamics of the lineage size $Z$ over one population life cycle is described by:

$$\frac{Z_i}{1-m} = 1 - w_i + \frac{w_i}{\bar{w}}\left(\sum_j \Pr(i \to j)Z_j + \Pr(i \to i)Z_i\right) \tag{4}$$

Here, the factor $1-m$ accounts for migration, invariably reducing the lineage size throughout its lifespan. The denominator $1-m$ represents the fraction culled in the present generation, while the last term in the right-hand side of (4) is a product of the fraction $w_i$ that survives and enters the next generation (hence divided by $\bar{w}$) and the sum of sizes of the daughter lineages descending from $i$ (Fig. 3B).

Eq (4) can be solved for $Z_i$ and the solution substituted into (3). The resulting equation makes it possible to express the mean fitness $\bar{w}$, a population characteristic, in terms of parameters that govern the dynamics within a single lineage. In a degenerative case where there is no recombination within the immigrant genotype, $\Pr(\pi \to j) = 0$, for all $j$, the lineage size can be found exactly:

$$Z_\pi = (1-m)\frac{\bar{w}(1-w_\pi)}{\bar{w} - w_\pi(1-m)} \tag{5}$$

and the only stable solution for the mean population fitness is $\bar{w} = 1 - m$, which is a standard result for one-locus migration-selection balance model (Nagylaki 1992).

**Frequencies of the foreign genotypes:** We now examine the equilibrium frequency dynamics of the intermediate length genotype $j$. Its frequency $p_j$ is decreased due to migration, selection and recombination into genotypes with fewer immigrant alleles, but increased due to recombination between genotypes with a greater number of immigrant alleles and the resident genotype (BARTON 1983). The frequency of the initial immigrant genotype $\pi$ can only be increased by migration at rate $m$. This leads to the following simple recursions:

$$\frac{p'_j}{1-m} = p_j \frac{w_j}{\bar{w}} \Pr(j \rightarrow j) + \sum_i p_i \frac{w_i}{\bar{w}} \Pr(i \rightarrow j) \tag{6}$$

where $\Pr(j \rightarrow j) = 1$ for a single gene, and

$$p'_\pi = (p_\pi + m) \frac{w_\pi}{\bar{w}} \Pr(\pi \rightarrow \pi) \tag{7}$$

for the initial immigrant type. At equilibrium, the corresponding frequencies $p^e$ are:

$$p^e_j = \frac{(1-m) \sum_i p_i w_i \Pr(i \rightarrow j)}{\bar{w} - (1-m) \Pr(j \rightarrow j) w_j} \tag{8}$$

and

$$p^e_\pi = \frac{m \Pr(\pi \rightarrow \pi)}{\Pr(j \rightarrow j) w_{in} - \bar{w}} \tag{9}$$

One can see that Eq. (8) and (9), which completely describe the genetic structure of the population at migration-selection balance, contain only those parameters that determine the branching process within a single lineage, except for the mean population fitness $\bar{w}$. It, in turn, can also be derived from the single lineage dynamics using Eq. (3). To obtain further results, we will make some simplifications of the probability function $\Pr(i \rightarrow j)$ that maps the transition between genotypes, and consider a specific model of constructing the fitness $w_i$ of the foreign genotype.

A SPECIFIC MODEL: INTROGRESSION OF LINEAR BLOCKS

**Multiple crossovers on a linear chromosome:** The need to follow the frequencies of a large number of genotypes, the number of which increases exponentially with the number of loci, is a major obstacle in multilocus population genetics. A popular solution is to lump the genotypes into classes that share the same number of alleles of a certain kind: we adopt this approach here as it is particularly useful in models of gene exchange between populations, where one is typically concerned with just two types of genes (i.e. the resident and the foreign alleles). Following Barton (1983), we assume that the foreign genes in the initial immigrant genotype are evenly distributed along a single chromosome, with a recombination rate $r$ between the two neighboring genes. This arrangement will be called a continuous *block* of genes (BAIRD 1995; BAIRD *et al.* 2003; UNGERER *et al.* 1998). The most important characteristic of the block is the number $k$ of the foreign genes it carries, this will be referred to as the *length* of the block; $k_\pi$ will represent the length of the initial immigrant block, $\pi$. Let us assume, for a start, that there can only be one crossing-over per block (BAIRD *et al.* 2003; BARTON 1983). In this case, the offspring genotypes will also be continuous blocks, and will only differ from the parental block by the number $k$ of the foreign alleles they inherit. The probability of getting an offspring block of any size in this case is simply $\Pr(i \to j) = 2r$ (Barton 1983). In the following, we depart from the single crossover assumption and allow for multiple independent recombination events per block. The probability of recombination occurring at $c$ particular locations chosen from $k-1$ intervals between $k$ genes is therefore $r^c (1-r)^{k-c-1}$. We then make another important simplification and treat the daughter genotypes resulting from multiple crossovers as continuous blocks, even though the actual map distance between the two neighboring foreign genes will no longer be uniform. That is, two genes next to each other will be assumed to recombine with the same probability $r$ as two genes separated by any number of the resident loci. We will demonstrate that although this assumption appears difficult to justify biologically, it almost always results in a more accurate estimation of population parameters than restricting the number of crossovers to one. We therefore only need to keep track on the lengths of the parental and daughter blocks, $k$ and $l$, and the probability of transition between full genotypes, $\Pr(i \rightarrow j)$, can be replaced by the probability of transition of the block from one length class to another:

$$\Pr(k \rightarrow l) = \sum_{c=1}^{k-1} r^c (1-r)^{k-c-1} \Psi(c,k,l) \tag{10}$$

where the function $\Psi(c,k,l)$ returns the number of daughter blocks of length $l$ that could be obtained from all combinations of $c$ crossovers on the parental block of length $k$ (see SI). It is given by:

$$\Psi(c,k,l) = \begin{bmatrix} \begin{pmatrix} l-1 \\ x \end{pmatrix} \begin{pmatrix} k-l-1 \\ x-1 \end{pmatrix} + \begin{pmatrix} l-1 \\ x-1 \end{pmatrix} \begin{pmatrix} k-l-1 \\ x \end{pmatrix}, \text{ if } c = 2x \\ 2 \begin{pmatrix} l-1 \\ x \end{pmatrix} \begin{pmatrix} k-l-1 \\ x \end{pmatrix}, \text{ if } c = 2x+1 \end{bmatrix} \tag{11}$$

We derive $\Psi(c,k,l)$ explicitly in the SI. Note that while there is exactly $\begin{pmatrix} k-1 \\ c \end{pmatrix}$ ways of choosing, with equal probability $r^c (1-r)^{k-c-1}$, the locations for $c$ crossovers among $k-1$ intervals between $k$ genes, only one pair of daughter blocks can result from the partition of a single parental block. This means that the function $\Psi(c,k,l)$ should also be divided on $\begin{pmatrix} k-1 \\ c \end{pmatrix}$, and the corresponding binomial cancels out in Eq (10).

**Fitness function with epistasis:** Approximating the multilocus genotypes as continuous gene blocks requires that the fitnesses of individual genotypes must be chosen such so they only depend on the number of foreign alleles and not on their positions. This is a reasonable assumption given a large degree of independence of gene function from the minor changes in its genomic localization. We will parameterize the fitness of a block of length $k$ as follows (Nick Barton, pers. comm.):

$$w_k = 1 - S + \frac{\left(1 - \dfrac{k}{n}\right)S}{1 - \dfrac{k}{n}\theta} \tag{12}$$

We use the ratio $\dfrac{k}{n}$, where $n$ is the arbitrarily chosen maximum number of loci that contribute to the local adaptation in the resident environment (= the length on the foreign chromosome in Barton (1983)), to establish a lower limit to the fitness of the worst possible genotype, which for $k = k_\pi = n$ is $w_\pi = 1 - S$, where $S$ is the selection acting on $n$ genes. At this point, the fitness is independent of the parameter of epistasis $\theta$. For any smaller block, $k < n$, however, the fitness is an increasing function of $\theta$, ($\theta < 1$), and at $\theta$ = 0, selection acts according to the additive scheme, $1 - \dfrac{k}{n}S$ (Fig. 4A). When negative, the parameter $\theta$ also imposes relatively stronger selection on smaller blocks than on larger blocks (positive epistasis). Positive values of $\theta$ in the feasible range of $0 < \theta < 1$ impose relatively stronger selection on the smaller blocks than on the larger blocks (negative epistasis, Fig. 4A).

Allowing for the initial block to carry only a fraction of the maximum number of selectable genes, $k_\pi < n$, makes it possible to compare the introgression of blocks of different lengths within the same parameter space. This is useful in the case where the populations involved in gene exchange have undergone incomplete divergence (i.e. the hypothetical diverging source population is still climbing the corresponding adaptive peak), as opposed to a scenario $k_\pi = n$ where further divergence is unlikely (the source population has settled on the peak). If $n$ is constant and $k_\pi$ is varied, selection pressure on the initial block increases with its length, whereas if $k_\pi = n$, and both are varied, the same selection pressure is distributed among different number of genes, that is, a polygenic trait can be compared to the phenotype controlled by only a few major loci.

Unless specified otherwise, the following results were obtained holding $n$ constant and varying $k_\pi$, $k_\pi < n$.

<center>NUMERICAL RESULTS: BRANCHING PROCESS AND SIMULATIONS</center>

The Eqs (3) and (8-9) can be solved numerically to obtain the characteristics of the population at migration-selection equilibrium, termed the Branching Process Approximation (BPA) in the following. Note that the existence of the equilibrium itself depends on the numerical values of the corresponding parameters, and even when the equilibrium exists, the branching process can adequately describe the population dynamics only when the foreign genotypes are rare. While finding the conditions of existence for the equilibria in the multilocus system is outside the scope of our paper, we established the validity of the numerical results by comparing them with the deterministic simulations where some or all of the BPA assumptions were lifted. The times to compute the numerical solutions are of orders of magnitude shorter than those required to perform the corresponding simulations: this allowed for a faster and deeper exploration of the parameter space. We concentrated on low values of the migration ($m < 10^{-2}$) and recombination ($r < 10^{-2}$) rates, and medium to strong selection ($S > 0.4$), since in this region of parameter space the introgression of the foreign genes is most likely to follow the branching process (see SI). Restricted recombination between loci that contribute to hybrid incompatibility has been inferred among many interbreeding taxa, with most obvious cases often resulting from the chromosomal rearrangement events (NOOR and BENNETT 2009).

The results of BPA were verified against the deterministic, frequency-based simulations of the migration-selection process: a detailed description of the simulations is provided in the SI. Overall, the branching process model was found to be a good approximation for the slow introgression at multiple loci (see SI, also BAIRD ET AL. 2003). Note that although Eq (3) for the mean population fitness has $k$ roots, corresponding to $k$ linear block genotypes, only the largest root is both stable and globally attractive, and was therefore used in the analysis.

**Extent of introgression and migration load:** Trivially, introgression of gene blocks is limited when selection is strong and recombination rate on a linear block is low. In our model, selection against gene blocks can be further strengthened or weakened by the epistatic interactions. In particular, increasing in the parameter of epistasis $\theta$ facilitated introgression and increased the migration load. We examined the effect of varying the length of the initial block $k_\pi$ in respect to the fixed maximum number of loci $n$.

Introgression of both very small and very large blocks resulted in lighter load, while blocks of intermediate length generally caused the heaviest load (Fig. 4B). This is due to the fact that both small and large blocks have limited opportunity of introgression into the resident gene pool: the small blocks penetrate the selection barrier easily, but can only produce short lineages, while the large blocks are confronted by a very strong barrier at the point when they are introduced into the native population. Since negative epistasis causes even stronger selective pressure on the large blocks, the convex shape of the load on Fig. 4B becomes more pronounced as $\theta$ increases. Note in Fig. 4A, that the full range of epistatic interactions depends on $n$ and not on $k_\pi$, reflecting the assumption that epistasis is determined by the adaptive landscape rather than the degree of divergence between populations. We used the replacements $n \to k_\pi$ and $S \to S\frac{k_\pi}{n}$ in the eq (12) to rescale both the epistasis and the strength of selection according to the length of the initial block (point 2 on Fig. 4A), but results were qualitatively similar to those presented on Fig. 4B (given in the SI). However, if the same selection pressure $S$ was distributed among different number of genes, that is, when the initial and the maximum block lengths were varied as a single parameter $k_\pi = n$, the migration load increased monotonically (Fig. 4C).

**Distribution of block frequencies and the effective selection on individual genes:** Irrespective of the value of the epistasis parameter, the frequency distribution of the block lengths is always bimodal, with the peaks corresponding to the smallest ($k = 1$) and the largest $\left(k = k_\pi\right)$ blocks (Fig. 5A). There is a gradual transition from the domination of the largest blocks under the strong positive epistasis ($\theta < 0$) to the domination of the single-gene blocks under the negative epistasis ($\theta > 0$, see Fig. 5A). For most part of its feasible range, the largest and the smallest block frequencies are of comparable scale: however, as $\theta$ approaches 1, the frequency of the single-gene blocks increases very rapidly: for example, at ($\theta = 0.8$, S = 0.7, $r = 0.01$ and $m = 0.001$) $p_1$ is an order of magnitude higher than $p_{15}$ (Fig. 5A). To demonstrate the extent of introgression by a single variable, we calculated the effective selection pressure on the individual genes, $s*$ (Barton 1983). This is simply the ratio of the migration rate, $m$, to the average weighted frequency of the single alleles at equilibrium, $\bar{p}$:

$$s^* = \frac{m}{\bar{p}} \tag{13}$$

where

$$\bar{p} = \sum_{k=1}^{k_\pi} \frac{w_k}{w_1} \frac{p_k}{k_\pi} \tag{14}$$

Note that $\bar{p}$ here accounts for the non-additive selection on the individual alleles, whereas in Barton (1983) only the additive fitness $\left(w_k = k w_1\right)$ was considered. The effective selection pressure roughly shows how much of the foreign genetic material is present in the resident population, relative to the number of migrants that arrive every generation. It takes the maximum value of 1 if there is no introgression (i.e. with positive migration rate, $m$, the fitness of the initial block, $w_\pi$, must be 0) and drops down to $m$ if gene flow swamps the resident population. We found that the effect of epistasis on the effective selection pressure is different from that on the distribution of block frequencies above. For the most part of the feasible range of $\theta$, $s*$ stays almost constant, but then drops abruptly when epistasis becomes strongly negative ($\theta >> 0$): this corresponds to the rapid increase in the frequency of the single gene blocks in the same parameter range (Fig. 5A, B). The change in $s$*, however, is much more gradual when the size of the initial block is less than the maximum allowed block length $\left(k_\pi < n\right)$, Fig. 5C).

<p style="text-align:center">INVASION OF THE MODIFIERS</p>

Since we are ignoring the gene positions on a linear block, it is not possible to estimate the strength of selection acting on the individual foreign genes. It is feasible, however, to follow the fate of the marker gene $x$ located at a considerable distance from the foreign block, such as that the recombination rate between the marker and the edge of the linear block, $r_x$, is much larger than the rate of recombination within the block itself $\left(r_x > r\right)$. The barrier to the gene flow at the neutral marker linked to a block of genes under selection has been calculated by Barton and Bengtsson (1986), by considering a matrix of all possible foreign backgrounds which the marker can be associated with. We use the same method here, but assume that the marker is a novel mutation equally likely to originate on both the foreign and the resident backgrounds. Instead of focusing on the barrier to gene flow, we investigate the fate of the mutation that reduces, to a certain degree, the deleterious effect of the foreign genes in a resident environment. In the fitness formulation used here, such mutation (modifier) can alter two key parameters in the Eq. 12: the strength of selection ($S$) and epistasis ($\theta$), by the corresponding replacements $S \to S_x$ and $\theta \to \theta_x$. We assume that the modifier's effect is manifested at the diploid stage, and since the foreign genes are found only in heterozygotes, the mutation is analogous to the modifier of dominance (KARLIN and MCGREGOR 1972). Note that in case the dominance at several loci needs to be modified at once: in principle, this is possible in polygenic systems with a few genes responsible for most of the variation in a trait and the remaining variation being affected by a much larger number of loci. For example, the distinctive wing coloration patterns in a butterfly *Heliconius erato* are controlled by three loci of major effect, expressed early in the wing development, while multiple QTLs shape the minor details during the later stages of development (PAPA ET AL. 2013). A mutation targeting genes early in the pathway would therefore interact with multiple genes expressed later on. Alternatively, the modifier can affect the recombination rate on its background genotype by the replacement $r \rightarrow \rho$, without changing the fitness of the foreign genotype (BARTON 1995). It follows from Eqs (8-10) that reduction of the recombination rate always limits the introgression of the foreign block into the resident population, thus improving the population mean fitness.

For the first time, we will consider the frequency $p_0 \neq 1$ of the resident gametes, calculated simply as $p_0 = 1 - \sum_{k=1}^{k_\pi} p_k$. This will not affect the frequencies of the foreign gene blocks, since they are only present in heterozygotes, but will require the mean population fitness being measured at the diploid stage. We calculate this simply as the frequency of heterozygotes formed by the foreign gametes after selection, $2\sum_{k=1}^{k_\pi} p_k w_k$, plus the fraction of the resident gamete pool that remains after all heterozygotes have been formed, $p_0 - \sum_{k=1}^{k_\pi} p_k$. This simplifies to:

$$\overline{w}^* = 1 - 2\sum_{k=1}^{k_\pi} p_k \left(1 + w_k\right) \tag{15}$$

With the resident haplotype is now included, there are $1 + k_\pi$ possible backgrounds that can carry the modifier: its frequencies on the $l$-th background indicated by $x_l$, $0 \leq l \leq k_\pi$, where 0 indicates the resident haplotype. In order to describe the dynamics of modifier genotypes, we have to follow both the transition of the marker between types and the transitions between foreign types themselves. The following recursions take place:

$$x_0' = \frac{\hat{x}_0 \hat{p}_0}{\overline{w}^*} + r_x \hat{p}_0 \sum_{k=1}^{k_\pi} \hat{x}_k \left(1 - r\right)^{l-1} \frac{w_{x,k}}{\overline{w}} + \left(1 - r_x\right)\hat{x}_0 \sum_{k=1}^{k_\pi} \hat{p}_k \left(1 - r\right)^{k-1} \frac{w_{x,k}}{\overline{w}^*} \tag{16}$$

for the modifier on the resident background, and

$$x_l' = (1-r)^{l-1} \frac{w_{x,l}}{\overline{w}^s} \left[ \hat{p}_0 \hat{x}_l (1-r_x) + \hat{p}_l \hat{x}_0 r_x \right] + \hat{p}_0 \sum_{k=l+1}^{k_\pi} \hat{x}_l \frac{w_{x,k}}{\overline{w}^s} \sum_{c=1}^{k-1} r^c (1-r)^{k-c-1} \left[ (1-r_x) \Psi_1(c,k,l) + r_x \Psi_2(c,k,l) \right]$$

(17)

for the modifier on the foreign background. Here, the «hat» superscript indicates the frequency of the corresponding type after migration took place, e.g. $\hat{x}_l = (1-m)x_l$ for all $l < k_\pi$ and $\hat{x}_l = (1-m)x_l + m$ for $l = k_\pi$; $w_{x,l}$ is the fitness of the foreign block associated with the modifier of selection or epistasis. In case where modifier only alters the recombination rate, $w_{x,k} = w_k$, $w_{x,l} = w_l$ and the background recombination rate $r$ is replaced by $\rho$ in the presence of the modifier. Differentiating the system (16 – 17) in respect to $x_l$ results in a matrix of linear coefficients: the leading eigenvalue of this matrix, $\lambda$, represents the strength of selection acting on the modifier (OTTO AND DAY 2011) while the population is still in the migration-selection equilibrium. As follows from the Eqs (16 – 17), the dynamics of the invading modifier is determined both by its own effect on the background genotype and by the population structure / parameters at the migration-selection equilibrium.

**Modifiers of selection, epistasis and recombination in a common parameter space:**
We systematically explored the parameter space in the Eqs (8 – 9) and (16 – 17) and numerically estimated the leading eigenvalue of the invasion matrix $\lambda$. Note that in our parameterization, the selective advantage due to the modifier mutation cannot exceed the fitness of the resident genotype: the modifier invasion rate $\lambda$ is therefore bounded by the difference between the (background) parameters of the migration-selection balance (i.e. $S$, $\theta$, $r$) and their corresponding altered values in the presence of the modifier $(\theta_x, S_x, \rho)$.

That is, the favorable conditions for the invasion of selection, epistasis and recombination modifiers are $S_x \ll S$, $\theta_x \ll \theta$ and $\rho \ll r$, respectively. The effects of background parameters on the invasion rates of all three modifier types are complex: we investigate these to some extent in the SI. In the following, we concentrate on the effect of the modifier's own properties, and arbitrary choose common set of background conditions that allow for the comparison between the different modifier types.

In general, the modifier of selection and epistasis are able to invade with the comparable rates under a wide range of conditions. The parametric plot on the Figure 6A shows that increasing the selection and epistasis differentials have similar effects on the invasion threshold $\lambda = 1$: the space where invasion is possible gets narrower as the linkage between the modifier and the foreign gene block ($r_x$) becomes loose (Fig. 6A). Even an unlinked modifier $\left( r_x = 0.5 \right)$ that simultaneously confers large selective advantage and switches the epistasis from positive to negative can invade the resident population. Reaching the invasion threshold for the unlinked modifier of recombination, however, requires much more stringent conditions, most of all a very high background recombination rate ($r \approx 0.2$). At these values of $r$, our model based on the branching process is no longer a good approximation to the multilocus introgression (Fig. S1, S2): we therefore assumed a tighter linkage to the foreign block ($r_x = 0.1$) to ensure that invasion is possible in the following examples. Throughout the parameter space, the invasion rate $\lambda$ is an exponentially increasing function of the altered epistasis parameter, $\theta_x$, and a linear function of the modifier selection parameter, $S_x$. In the example shown on Fig. 6B, the maximum strength of selection acting on the modifier approaches (but can never exceed) the migration load. The background parameter set on Fig. 6B intersects with that on Fig. 6C, where $\lambda$ is plotted as a function of the altered recombination rate $\rho$. Although comparable, the invasion rate of the recombination modifier is much less than that of selection/epistasis, and the threshold for invasion is only reached if the recombination on a block is heavily suppressed or completely arrested ($\rho \approx 0$). Note that due to our model being a good approximation only when the linkage between foreign loci is sufficiently tight, we could not compare the case where the modifier of recombination is expected to spread with the fastest rate, i.e. where the recombination between the unlinked immigrant genes is arrested by the modifier. To find out if the modifier arresting recombination (from the value of $r = 0.05$) on the larger blocks is favored over that on the smaller blocks, we set $\rho = 0$ and estimated $\lambda$ against $k_\pi$ under the varying background conditions. The invasion rate was indeed increasing with the initial block length, though still being of order of magnitude smaller than the corresponding migration load (Fig. 6D).

Under the negative epistasis ($\theta > 0$) the modifier arresting recombination on very large blocks ($k_\pi > 10$) invades marginally slower than that on the blocks of moderate length: this parallels reduction in migration load imposed by large initial blocks (Fig. 4, Fig. 6D).

<div align="center">DISCUSSION</div>

In this article, we approximated introgression at multiple loci based on the assumption that the foreign genotypes occupy a very small proportion of the resident population and that the migration-selection process is at equilibrium. First, a general analytical description of introgression as a multi-type branching process was derived, based on the transition probability between recombining genotypes; we then simplified it further by assuming a specific model of gene blocks, where the population mean fitness and the genotype frequencies could be found numerically. Finally, we demonstrated how these equilibrium results could be used to investigate the invasion of a modifier mutation that masks the deleterious effect of gene flow, or suppresses recombination between the immigrant genes. Bearing in mind that our model only provides a good approximation at the low migration, low recombination scenario, we found wide parameter ranges where selection for the masking modifier was significant relative to the migration load imposed on the resident population, but narrower range of feasible conditions allowed for invasion of the recombination modifier.

**Comparison to Barton (1983):** Modeling introgression as a branching process where the linear gene blocks, explicitly characterized by the number of genes they carry, are being broken by recombination goes back to Barton (1983). He treated the resident population as (i) an unlimited pool of mating partners, which (ii) has essentially no effect on the dynamics of branching lineages. In the present paper, we kept the first assumption, but lifted the second one: that is, the foreign genotypes were still allowed to mate freely, and exclusively, with the residents, but selection against the migrants was normalized by the mean fitness of the resident population, $\bar{w}$. While Barton's study mostly concerned the balance between selection and recombination rates, which ultimately determined whether introgression is limited or indefinite, we concentrated on the effect of the (limited) introgression on the resident population structure. Unless the immigrant lineages can be treated as independent, valid approximation of this effect by the branching process is not possible: this is why our numerical results are presented for such parameter ranges (low migration and recombination rates, and strong selection) that keep the total fraction of the introgressed genetic material low at equilibrium. Other distinctive feature of our model is that multiple crossovers per gene block are allowed, rather then just a single crossover (BAIRD *et al.* 2003; BARTON 1983). While our approach leads to correct (binomial-like) distribution of the daughter block lengths following recombination of the initial immigrant block of $k_\pi$ genes, it underestimates the recombination rate at the discontinuous daughter blocks that contain «gaps» between the foreign genes. The single crossover model, however, gives an incorrect (uniform) distribution of the daughter block lengths, and ignores the discontinuous blocks altogether. The position-independent model, and the single crossover model, respectively, under- and overestimate the frequencies of the small blocks, but the differences between the two models are small as long as the recombination rates are low (Fig. S2).

In our model, selection on different genotypes within the immigrant lineage depended significantly on the epistatic interations within a gene block. Yet, the effect of introgression estimated on the population level (migration load, Fig. 4) and on the level of individual genes (the effective selection pressure *s**, Fig. 5B,C), revealed that epistasis only has a strong overall effect if the deviation from additive fitness extends to the initial immigrant genotype. If, however, the initial genotype is insensitive to epistasis, as in the case of $k_\pi = n$, the consequences of fitness non-additivity are much more subtle, and are visible primarily on the shape of the block frequency distribution (Fig. 5A). This is because the smaller blocks represented at higher frequency under the negative epistasis experience relatively weaker selection than they do under the posisitive epistasis, despite being less abundant in the latter case. Our results thus suggest that under a specific assumption that epistasis is manifested in the offspring of the initial block but has little of no effect on the initial block itself, the same amount of load and per-gene effective selection can be generated with very different observable patterns of introgression.

**Alternative microevolutionary processes at the migration-selection balance:** In the presence of the maladaptive gene flow, selection should favor those resident genotypes that have less chance of being associated with the immigrant genes. An allele that contributes to a stronger reproductive barrier between the residents and the migrants, through genetic, ecological, behavioral or other incompatibility, should therefore invade: current theory and empirical evidence suggest that that reinforcement or prezygotic isolation can occur in many systems (ABBOTT *et al.* 2013; SERVEDIO and NOOR 2003), though it is not particularly likely in a one-way migration scenario (SERVEDIO AND KIRKPATRICK 1997). Alternatively, a rare variant, which improves fitness while being found with the immigrant allele, relative to the most widespread type in the population, can also be favored: whether this mechanism operates in nature remains to be demonstrated, but models of the evolution of dominance in spatially heterogeneous environment (OTTO AND BOURGUET 1999; PROULX AND PHILLIPS 2005) and invasion of gene duplication in response to gene flow (YANCHUKOV AND PROULX 2012) suggest that it is at least likely under a wide range of conditions. Moreover, an analysis of fitness gradients for mate choice versus the modification of dominance (DURINX AND VAN DOOREN 2009) suggested that the latter is more strongly selected when allele frequencies at the single locus with heterozygote disadvantage are unequal: the situation where rare migrants are introduced into a large resident population satisfies this criterion. Here, we showed that robustness to the deleterious effect of immigrant alleles in a multi-locus system can evolve at the migration-selection balance, but the question of whether this scenario poses a likely alternative to the reinforcement of the reproductive barriers remains open. A direct comparison of selection strengths acting on the modifier(s) of assortment (PROULX AND SERVEDIO 2009) with the modifiers of selection/dominance/recombination is therefore needed: a study on this subject is currently under way.

**Evolution of recombination rate:** In a multilocus system, a strong barrier to gene flow exists even in the absence of any mechanisms of pre-mating isolation (BARTON AND BENGTSSON 1986). This is due to the fact that recombination is generally a much slower process than selective elimination of individuals carrying the maladaptive foreign alleles in linkage disequilibrium: physically reducing the rate of recombination between the foreign genes would therefore make the barrier even stronger. A number of models of migration-selection balance suggest that suppression of recombination / clustering of the locally adaptive loci in the genome is favored by selection (BANK *et al.* 2012; LENORMAND and OTTO 2000; YEAMAN and OTTO 2011; YEAMAN and WHITLOCK 2011). Here, we have demonstrated that tightening linkage between the immigrant genes increases the population mean fitness (by reducing the migration load) and therefore the modifier that reduces the recombination rate is favored by selection (Fig. 6C,D). However, under the premise that the linkage on the introgressing block is already tight, further reduction, or even complete arrest of recombination cannot be selected too strongly. This caveat could explain why a direct comparison between the invasion rates of the modifiers of selection/epistasis and the modifier of recombination revealed much weaker selection for the latter (Fig. 6B,C).

**Effect of the initial block length**: Despite the fact that, after recombining into the resident gene pool, the foreign blocks of different length are selected independently at any single time point, selection on the larger blocks negatively affects the smaller blocks in a descending lineage. Increasing the number of genes in the initial block has therefore a dual effect on the lineage size: even though maladapted genes carried by large chunks of the genome are inherited by a larger number of descendants, elimination of the large blocks early in the lineage can be so effective that their impact on the equilibrium population fitness, measured as the migration load, becomes less relative to that caused by the introgression of the blocks carrying moderate number of genes (Fig. 4). This non-monotonic effect on the migration load is also manifested on the invasion rates of various modifier types, though with much weaker intensity (Fig. 6D, Fig. S3). While a modifier mutation rescuing a (single) large block has a stronger selective advantage, it can still have less chance of being associated with the foreign alleles if the introgression is limited by the initial block size. Note that the standard matrix method of calculating the invasion rate employed here does not distinguish between the modifier originating on any specific genetic background. An alternative method would be to calculate the speed of the branching process starting from the modifier mutation associated *with the particular* *genotype*: this would also be the only feasible approach of examining the modifier which is tightly linked to the foreign block. So far, our results only seem to suggest that the gene flow dynamics in heterogeneous populations may depend non-linearly on the size of the genomic region containing the locally adapted loci.

## Acknowledgements

This study was preceded by the discussion with Nick Barton of the ways to calculate the migration load, we appreciate his valuable suggestions therein. We are also grateful to Alex Ross for the derivation of Eq. 11 and to Jay Taylor, Joao Hespahna and Kyoungmin Roh for the discussion of our preliminary results. This work has been supported by the NSF grant EF-0742582 to SP.

LITERATURE CITED

ABBOTT, R., D. ALBACH, S. ANSELL, J. W. ARNTZEN, S. J. BAIRD *et al.*, 2013 Hybridization and speciation. J Evol Biol **26:** 229-246.

AGRAWAL, A., and M. WHITLOCK, 2012 Mutation Load: The Fitness of Individuals in Populations Where Deleterious Alleles Are Abundant. Annual Review of Ecology, Evolution, and Systematics **Vol. 43:** 115-135.

BAIRD, S. J., 1995 A simulation study of multilocus clines. Evolution **49:** 1038-1045.

BAIRD, S. J., N. H. BARTON and A. M. ETHERIDGE, 2003 The distribution of surviving blocks of an ancestral genome. Theor Popul Biol **64:** 451-471.

BANK, C., R. BURGER and J. HERMISSON, 2012 The limits to parapatric speciation: Dobzhansky-Muller incompatibilities in a continent-island model. Genetics **191:** 845-863.

BARTON, N. H., 1983 Multilocus clines. Evolution **37:** 454-471.

BARTON, N. H., 1995 A general model for the evolution of recombination. Genet Res **65:** 123-145.

BARTON, N. H., 2002 Multilocus Package. http://www.biology.ed.ac.uk/research/groups/barton/, pp.

BARTON, N. H., and B. O. BENGTSSON, 1986 The barrier to genetic exchange between hybridising populations. Heredity **57:** 357–376.

BILLIARD, S., and T. LENORMAND, 2005 Evolution of migration under kin selection and local adaptation. Evolution **59:** 13-23.

BLANQUART, F., S. GANDON and S. L. NUISMER, 2012 The effects of migration and drift on local adaptation to a heterogeneous environment. J Evol Biol **25:** 1351-1363.

BOLNICK, D. I., and P. NOSIL, 2007 Natural selection in populations subject to a migration load. Evolution **61:** 2229-2243.

DURINX, M., and T. J. VAN DOOREN, 2009 Assortative mate choice and dominance modification: alternative ways of removing heterozygote disadvantage. Evolution **63:** 334-352.

GRAY, J. C., and M. R. GODDARD, 2012 Gene-flow between niches facilitates local adaptation in sexual populations. Ecology Letters **15:** 955-962.

HALDANE, J. B. S., 1937 The effect of variation on fitness. American Naturalist **71:** 337-349.

HALDANE, J. B. S., 1957 The cost of natural selection. Journal of Genetics **55:** 511-524.

HARRISON, R. G., 1993 *Hybrid zones and the evolutionary process.* Oxford University Press on Demand.

HEATHCOTE, C. R., 1965 A Branching Process Allowing Immigration. Journal of the Royal Statistical Society **27:** 138-143.

JIGGINS, C. D., and J. MALLET, 2000 Bimodal hybrid zones and speciation. Trends Ecol Evol **15:** 250-255.

KARLIN, S., and J. MCGREGOR, 1972 Towards a theory of the evolution of modifier genes. Theoretical Population Biology **5:** 59-103.

LENORMAND, T., and S. P. OTTO, 2000 The evolution of recombination in a heterogeneous environment. Genetics **156:** 423-438.

NADEAU, N. J., A. WHIBLEY, R. T. JONES, J. W. DAVEY, K. K. DASMAHAPATRA *et al.*, 2012 Genomic islands of divergence in hybridizing Heliconius butterflies identified by large-scale targeted sequencing. Philos Trans R Soc Lond B Biol Sci **367:** 343-353.

NAGYLAKI, T., 1992 *Introduction to Theoretical Population Genetics*. Springer.

NOOR, M. A., and S. M. BENNETT, 2009 Islands of speciation or mirages in the desert? Examining the role of restricted recombination in maintaining species. Heredity (Edinb) **103:** 439-444.

OTTO, S. P., and D. BOURGUET, 1999 Balanced Polymorphisms and the Evolution of Dominance. The American Naturalist **153:** 561-574.

OTTO, S. P., and T. DAY, 2011 *A Biologist's Guide to Mathematical Modeling in Ecology and Evolution*. Princeton University Press.

PAPA, R., D. D. KAPAN, B. A. COUNTERMAN, K. MALDONADO, D. P. LINDSTROM *et al.*, 2013 Multi-allelic major effect genes interact with minor effect QTLs to control adaptive color pattern variation in Heliconius erato. PLoS One **8:** e57033.

PROULX, S. R., 2001 Female choice via indicator traits easily evolves in the face of recombination and migration. Evolution **55:** 2401-2411.

PROULX, S. R., and P. C. PHILLIPS, 2005 The opportunity for canalization and the evolution of genetic networks. Am Nat **165:** 147-162.

PROULX, S. R., and M. R. SERVEDIO, 2009 Dissecting selection on female mating preferences during secondary contact. Evolution **63:** 2031-2046.

RIESEBERG, L. H., J. WHITTON and K. GARDNER, 1999 Hybrid zones and the genetic architecture of a barrier to gene flow between two sunflower species. Genetics **152:** 713-727.

RONCE, O., and M. KIRKPATRICK, 2001 When sources become sinks: migrational meltdown in heterogeneous habitats. Evolution **55:** 1520-1531.

SEEHAUSEN, O., 2004 Hybridization and adaptive radiation. Trends Ecol Evol **19:** 198-207.

SERVEDIO, M. R., 2007 Male versus female mate choice: sexual selection and the evolution of species recognition via reinforcement. Evolution **61:** 2772-2789.

SERVEDIO, M. R., and M. KIRKPATRICK, 1997 The Effects of Gene Flow on Reinforcement. Evolution**:** 1764-1772.

SERVEDIO, M. R., and M. NOOR, 2003 The role of reinforcement in speciation: theory and data. Annual Review of Ecology, Evolution, and Systematics**:** 339-364.

STAR, B., R. J. STOFFELS and H. G. SPENCER, 2007 Single-locus polymorphism in a heterogeneous two-deme model. Genetics **176:** 1625-1633.

STORFER, A., J. CROSS, V. RUSH and J. CARUSO, 1999 Adaptive coloration and gene flow as a constraint to local adaptation in the streamside salamander, Ambystoma barbouri. Evolution**:** 889-898.

SZYMURA, J. M., and N. H. BARTON, 1991 The genetic structure of the hybrid zone between the fire-bellied toads Bombina bombina and B. variegata: comparisons between transects and between loci. Evolution**:** 237-261.

TURNER, T. L., M. W. HAHN and S. V. NUZHDIN, 2005 Genomic islands of speciation in Anopheles gambiae. PLoS Biol **3:** e285.

Ungerer, M. C., S. J. Baird, J. Pan and L. H. Rieseberg, 1998 Rapid hybrid speciation in wild sunflowers. Proc Natl Acad Sci U S A **95:** 11757-11762.

van Doorn, G. S., and F. J. Weissing, 2006 Sexual conflict and the evolution of female preferences for indicators of male quality. Am Nat **168:** 742-757.

Yanchukov, A., and S. R. Proulx, 2012 Invasion of gene duplication through masking for maladaptive gene flow. Evolution **66:** 1543-1555.

Yeaman, S., and S. P. Otto, 2011 Establishment and maintenance of adaptive genetic divergence under migration, selection, and drift. Evolution **65:** 2123-2129.

Yeaman, S., and M. C. Whitlock, 2011 The genetic architecture of adaptation under migration-selection balance. Evolution **65:** 1897-1911.

**Figure Legends**

**Figure 1**. A subset of the lineage descending from an immigrant haplotype $\pi$ carrying 10 genes. After selection, the genotypes carrying foreign genes (solid black dots) recombine with the resident genotypes (shown as the rows of open dots), so that the average number of foreign alleles per genotype is reduced every generation $(t_{0-3})$. Once recombination has broken the foreign genotype down to one gene, it is certain to eventually be eliminated by selection in the single foreign allele state.

**Figure 2**. Migration load at the migration-selection balance. The reduction in mean population fitness caused by the descendants of the migrants arrived in the generation 20 is measured every generation that follows (large black dots); this is equivalent to the load caused by the progeny of the migrants arrived in all preceding generations, calculated at point $t = 20$ (smaller black dots).

**Figure 3**. **A** – Schematic representation of the terms in Eq (3) for mean population fitness. The future load caused by the independent foreign lineages (white contour) is subtracted from the fitness contribution of the migrants, measured at the generation of sampling (contour filled by dark grey). Note that the proportion of migrants culled before reproduction, $m(1-w_\pi)$, is not a part of the mean population fitness. **B** – Dynamics of the lineage size, $Z_i$, determined by the Eq. (4). $Z_i$ is reduced by migration and selection, but increased as the foreign genotype recombines and gives rise to sub-lineages. Note that the term $(1-m)(1-w_i)$, i.e. the contribution of the individuals culled by selection, is included in the lineage size calculation.

**Figure 4.** Migration load and the epistatic interaction between genes on a linear block. **A** – The fitness of the genotype $w_k$ is plotted against the proportion of the chromosome $\dfrac{k}{n}$ occupied by the foreign genes: $\theta = 0$ corresponds to additive fitness (no epistasis), while $\theta = -5$ and $\theta = 0.8$ represent the positive and the negative epistasis, correspondingly. The points indicate the fitness of the initial block carrying 4 genes $(k_\pi = 4)$ at $\theta = 0.8$. Point 1: $n = 10$, fitness is calculated according to eq. 12. Point 2: $n = 10$, fitness is rescaled by the length of the initial block. Point 3: initial block carries the maximum number of genes, $k_\pi = n = 4$. **B** – Migration load ($L$), calculated using the Branching Process Approximation (BPA), as a function of the length ($k_\pi$) of the initial block. The curves (from bottom to top) correspond to $\theta =$ -10, -3, -1, 0, 0.5, 0.8. Parameter values are: $m =$ 0.003; $r = 0.015$; $S = 0.7$; **C** – the same as in B, but $n = k_\pi$.

**Figure 5.** **A** – distribution of the foreign genotype frequencies at equilibrium, with the initial block composed of 15 genes, obtained through the Branching Process Approximation (BPA) by solving the Eqs (3) and (8-9) numerically. Note that the actual frequency of the single-gene blocks at $\theta =0.8$ is much higher than the corresponding bar height on the plot (indicated by the larger font on the Y-axis). **B** – effective selection, $s^*$, as a proportion of the total selection on the initial block, with the number of genes in the initial block varying from 1 to 15, and epistasis ranging from -9 to 0.8. The maximum number of genes is fixed at $n = 15$. **C** – the same as in B, but the maximum number of genes, $n$, is set to equal $k_\pi$ in each case. Points along the curve for 15 loci correspond to the frequency distribution plots on A. Other parameters: $S = 0.7$, $r = 0.01$, $m = 0.001$.

**Figure 6.** Selection acting on the invading modifier. **A** – view of the parameter space with the modifier selection strength $(S_x)$ and epistasis $(\theta_x)$ plotted on the X and Y axes, the leading eigenvalue of the invasion matrix (λ) equals one along the lines shown for the following values of $r_x$ (top to bottom): 0.5, 0.3, 0.1, 0.05. The modifier is able to invade in the space above each line, and cannot invade below the line(s). Other parameters: $S = 0.7$, $\theta = 0.2$, $m = 0.003$, $r = 0.001$, $k_\pi = 5$. **B** – λ as a function of the modifier epistasis, $(\theta_x)$, plotted on the bottom axis; and selection strength $(S_x)$, plotted on the top axis. The migration load (which is independent of $S_x, \theta_x$) is shown by the horizontal line on the top. Solid lines indicate $k_\pi = 5$, dotted lines indicate $k_\pi = 7$. Other parameters: $S = 0.7$, $\theta = -6$,

$m = 0.001$, $r = 0.05$, $r_x = 0.1$, $n = 10$. **C** – Invasion rate of the modifier of recombination, as a function of the imposed recombination rate $\rho$. Markers indicate different values of the background epistasis, solid lines: $k_\pi = 5$, dotted lines: $k_\pi = 7$. Other parameters are as in B. **D** – Modifier of recombination invades non-monotonically with the length of the initial block, $k_\pi$. Solid lines: $k_\pi = 5$, dotted lines: $k_\pi = 7$; markers indicate different values of the background epistasis. Migration load is shown at the top of the plot. Other parameters: $S = 0.7$, $m = 0.003$, $r = 0.05$, $r_x = 0.1$, $n = 12$.

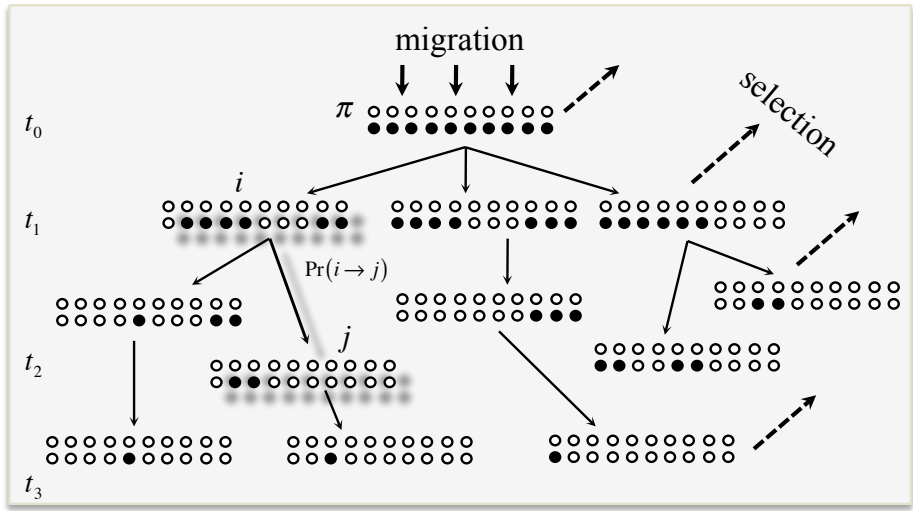

**Figure 1.**

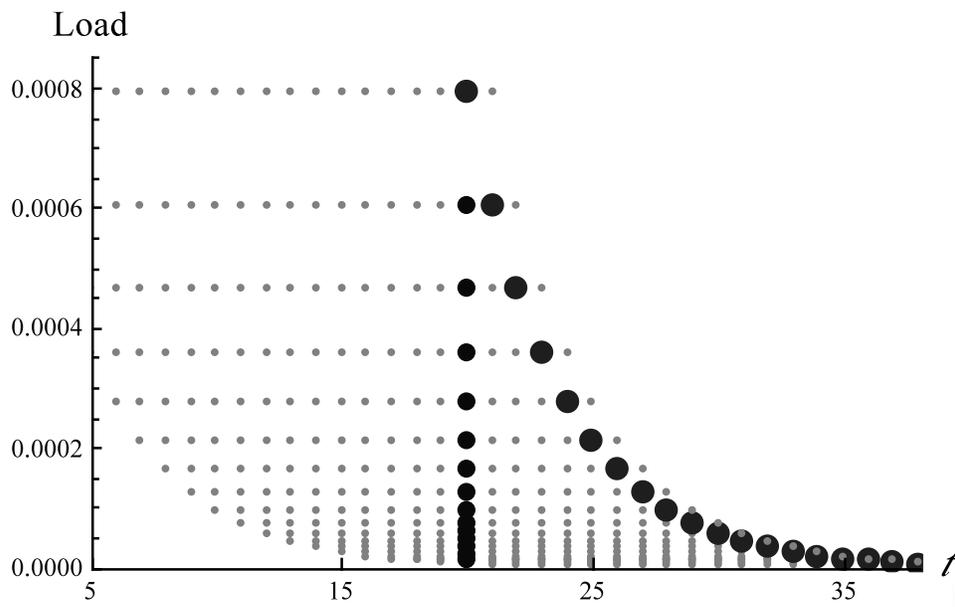

**Figure 2.**

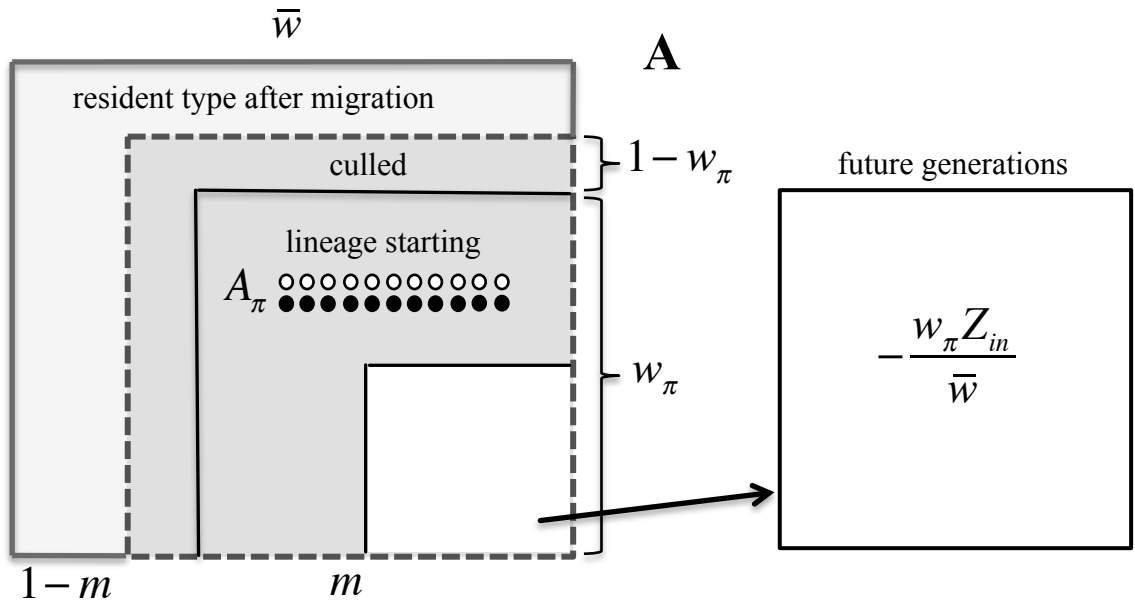

**Figure 3A**

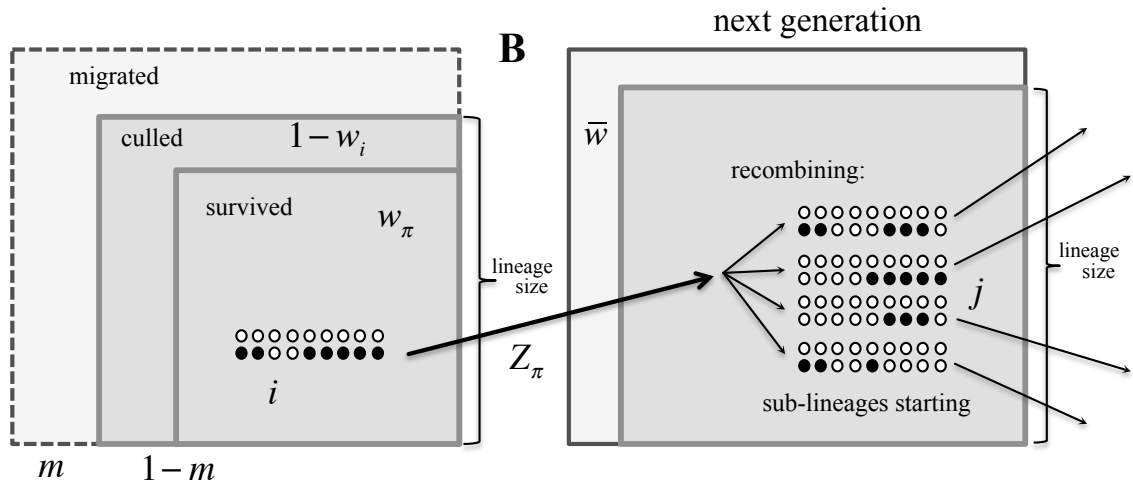

**Figure 3B**

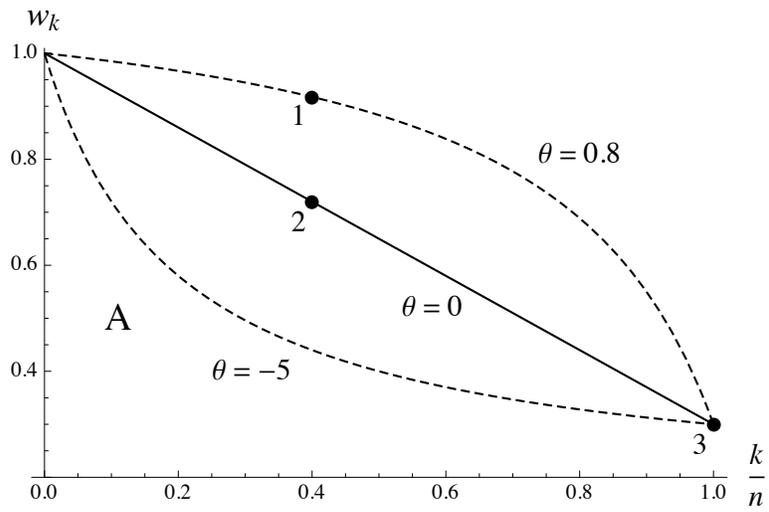

**Figure 4A**

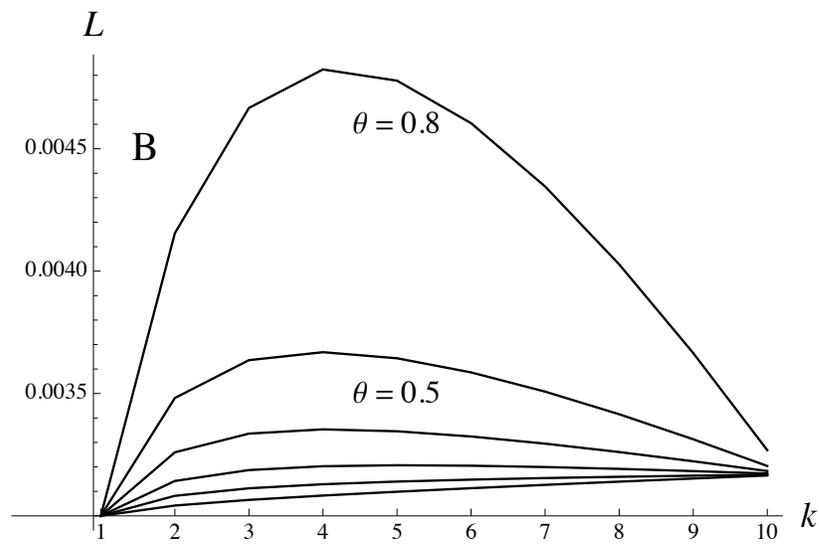

**Figure 4B**

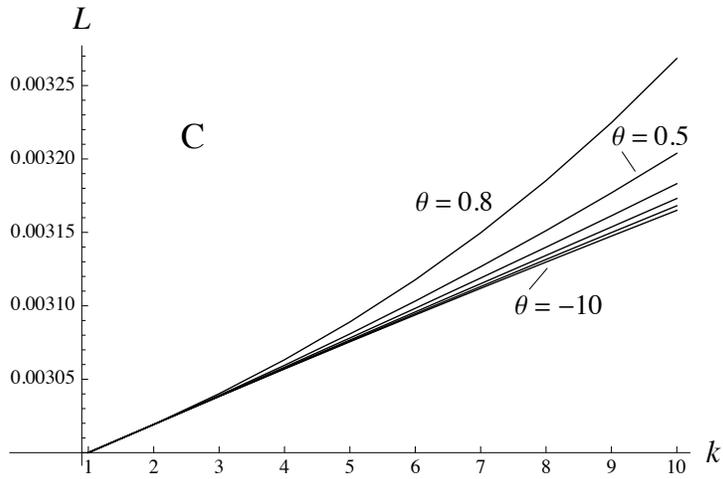

**Figure 4C**

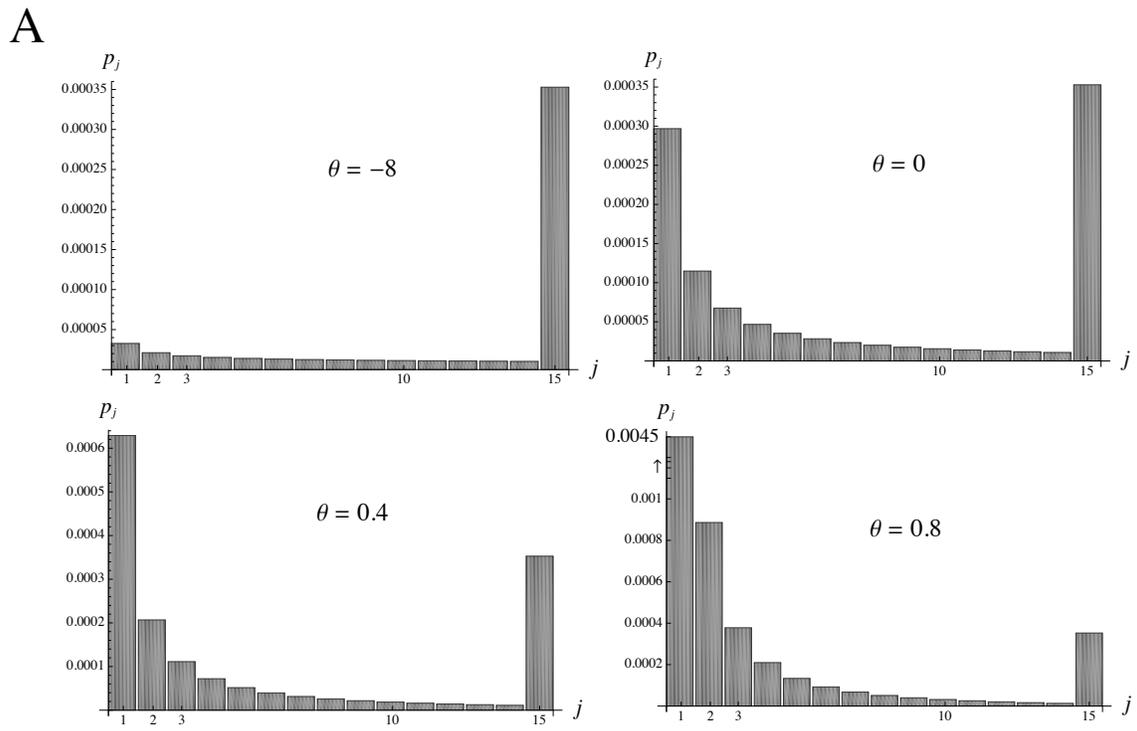

**Figure 5A**

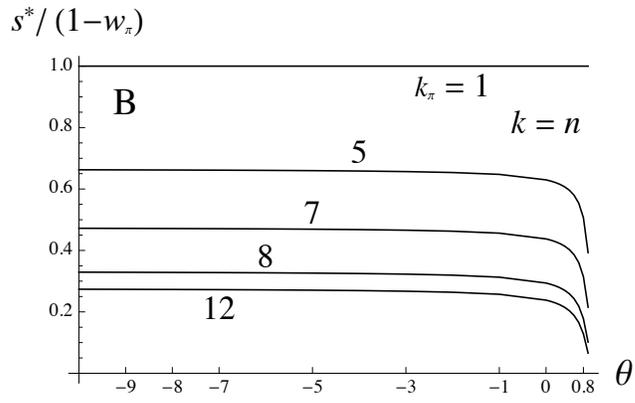

**Figure 5B**

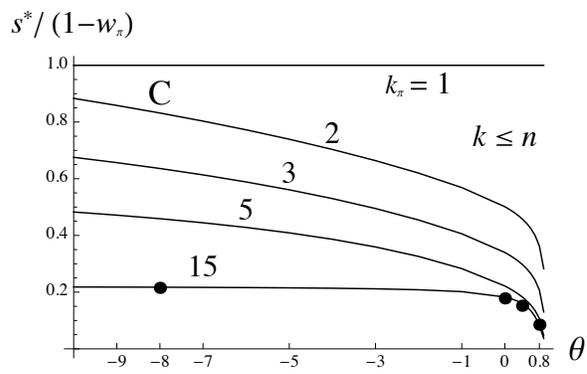

**Figure 5C**

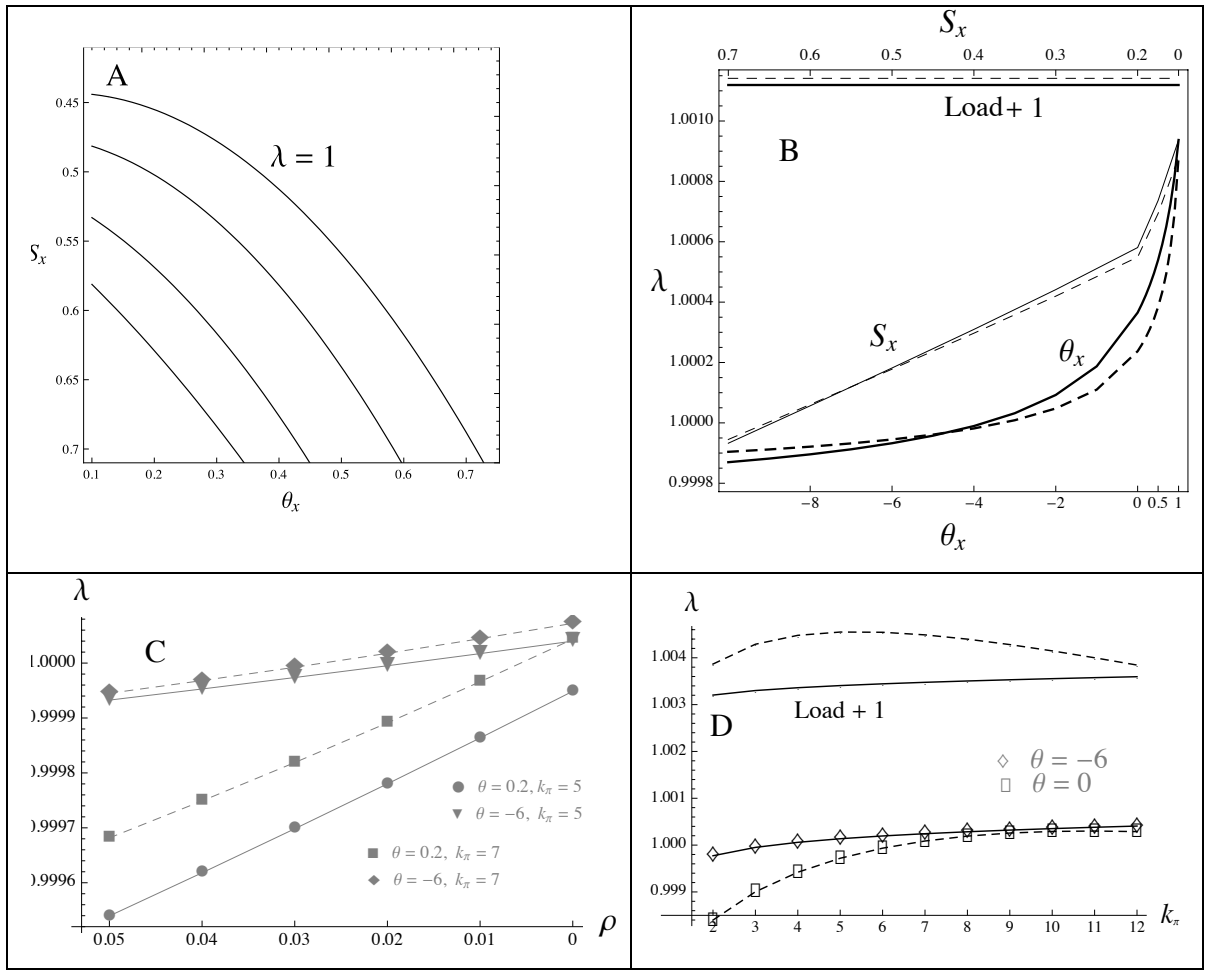

**Figure 6**

SUPPORTING INFORMATION

**I. Testing the results of the Branching Process Approximation against simulations**

We verified the validity of Branching Process Approximation against the deterministic, frequency-based simulations of the migration-selection process as it was described in the previous sections. All simulations started from the population consisting of the resident genotypes only; the foreign genotypes, represented as identical linear blocks of equally spaced genes, were introduced every generation at the rate $m$, followed by selection and reproduction. The simulations were run using the Multilocus package (Barton 2002) in *Mathematica* until the equilibrium was reached, at which point the stable genotype frequencies were compared with the corresponding results of BPA. Two simulation models were used. In the Introgressive Simulation (IS), the foreign genotypes could only mate with the resident genotypes, assuming that there an unlimited supply of the residents. At the stage of selection, however, the frequency of the resident genotype was calculated as $p_0 = 1 - \sum_j p_j$, and the mean population fitness as $p_0 + \sum_j p_j w_j$. The IS therefore exactly follows Eqs (3) and (6-7), which describe a general branching process, without the simplifying assumption of all foreign genotypes being represented as continuous gene blocks. If the initial genotype consists of only one ($k_\pi = 1$) or two ($k_\pi = 2$) genes, the results of the introgressive simulation match the BPA precisely (Fig. S2 in SI), but starting from $k_\pi = 3$, the IS includes genotypes that carry the same number of genes located at different positions. The deviation of the IS from BPA increases as the number of genes, migration and recombination rates increase. As a modification of the IS, we included the case where the number of crossovers per block was restricted to one, referred to as the Single Crossover Simulation (SCS) in the following. The probability of recombination on a block of length $k$ was calculated in SCS as $(k-1)r$, which corresponds exactly to the recombination model used by Barton (1983).

In the Panmictic Simulation (PS), all mating types, including those between foreign genotypes, were allowed, and the positions of individual genes were recorded in all genotypes. The PS is the most realistic way to describe the population at the migration-selection balance, following the frequencies of $2^k$ genotypes composed of two (the resident and the foreign) alleles at $k$ loci. Selection in PS was acting on haploids, consistent with the selection regime in BPA and IS. Since matings between the foreign genotypes were still very rare, a switch to diploid selection (acting differently on zygotes that carry foreign genes on both homologous sets) had very little effect on the results. As expected, the PS shows larger deviation from BPA than the Introgressive Simulation, with the differences between PS and IS typically being greater than between IS and BPA (Fig. S1, S2). The SCS, however, consistently deviated most from the PS in comparison to the IS and BPA. Nevertheless, the deviation remains small in the parameter range where our numerical results were obtained (Fig. S1, S2), suggesting that the branching process model is a good approximation for the introgression at multiple loci (Baird et al. 2003).

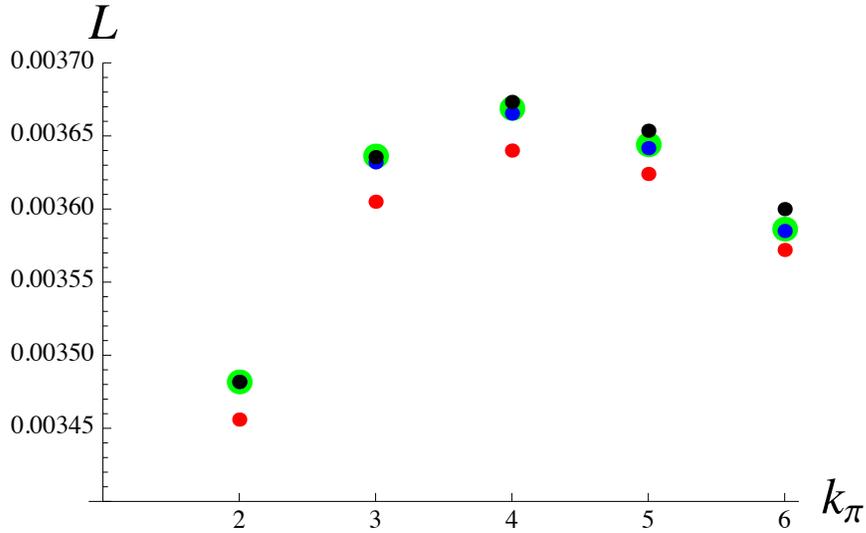

Figure S1. Migration load imposed by the introgression of 2 to 6 genes in the initial block, calculated in BPA (green dots), Panmictic Simulation (PS, red dots), Introgressive Simulation (IS, blue dots) and Single Crossover Simulation (SCS, black dots). Parameter values are: $S = 0.7$, $\theta = 0.5$, $m = 0.003$, $r = 0.015$, $n = 10$.

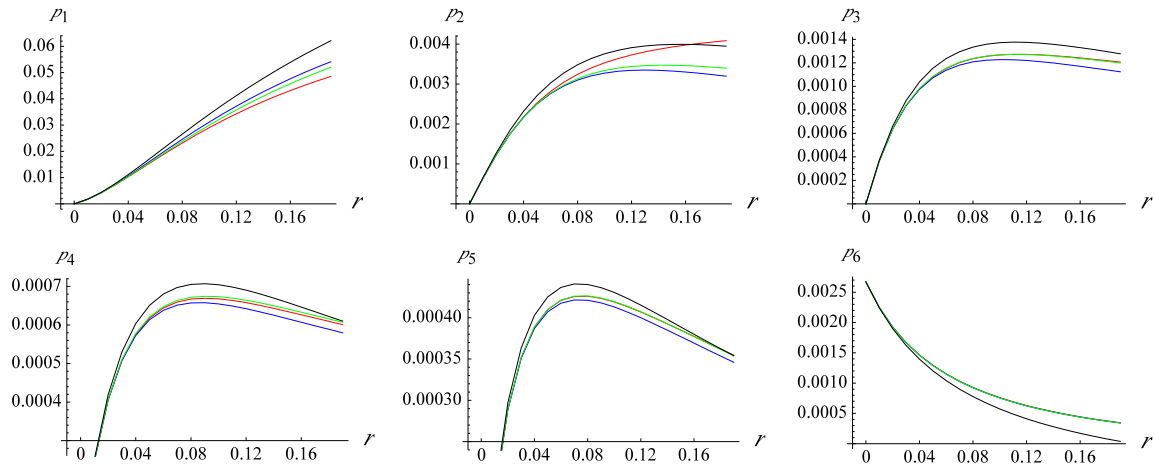

Figure S2. Comparison of the foreign genotype frequency distributions, with the initial block composed of 6 genes, obtained through BPA (green line) against Panmictic Simulation (PS, red line), Introgressive Simulation (IS, blue line) and Single Crossover

Simulation (SCS, black line) as the recombination rate ($r$) increases. For the simulation results to be comparable with the BPA, different genotypes were grouped together by the number of genes ($j$) they carry. Each plot represents variation in the frequency of the corresponding gene block (1-6). Parameter values are: $S = 0.5$, $\theta = 0.2$, $m = 0.001$.

## II. Effect of the background parameters and population structure

Increasing the background selection strength on the foreign chromosome, $S$, has a strong positive effect on the invasion rate, at the same time reducing the gap between the load and $\lambda$ (Fig. S3A). When $\theta$ and $\theta_x$ are fixed, the modifier associated with the larger initial block ($k_\pi = 5$) invades faster relative to the one associated with the smaller block ($k_\pi = 3$). Varying the background epistasis parameter $\theta$, however, has little effect while the epistasis is positive $(\theta < 0)$, but results in an exponential drop in $\lambda$ once the epistasis becomes negative $(\theta > 0)$, see Fig. S3A. When $S$ and $S_x$ are fixed, the effect of the number of genes in the initial block is reversed: the larger blocks inhibit the invasion of the modifier while the smaller blocks increase the leading eigenvalue (Fig. S3B). Note that negative epistasis also increases the total frequency of the foreign genetic material in the population and, therefore, migration load (Fig. 4, Fig. 5 in the paper). To separate the effect of the population structure from the difference between the background and the altered epistasis parameters, we equalized and simultaneously varied both $(\theta = \theta_x)$. There was little variation in $\lambda$ for most of the parameters range, but then a rapid non-linear response once the epistasis became extremely negative (Fig. S3C). Such response is greatly alleviated by the modifier's distance $r_x$ from the foreign block: note that our model is feasible for only those values of $r_x$ that are much larger than the recombination rate on the block itself. Opposite to the negative effect of $r_x$, the background recombination rate on the foreign block, $r$, is positively and almost linearly correlated with selection on the invading modifier. This is because the large blocks are broken down faster as *r* increases, creating more opportunities for the modifier to be picked up by selection, which in turn results in positive correlation between $\lambda$ and *r* (Fig. S3D).

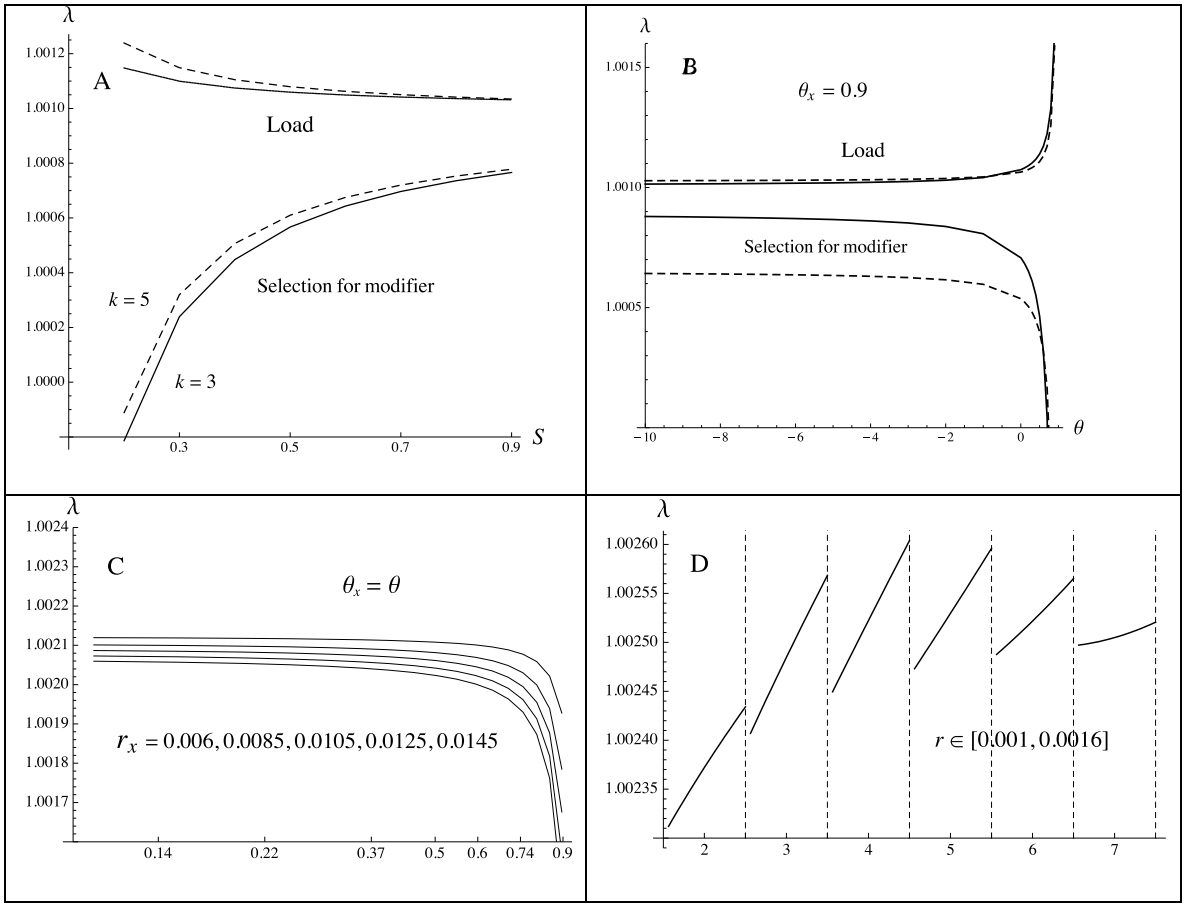

Figure S3. **A** – effect of the background selection $(S)$, shown next to the load. Solid and doted lines indicate that the initial block is composed of $k = 3$ and $k = 5$ genes, respectively. $\theta = 0.3$, $m = 0.003$, $r = 0.01$, $S_x = 0.2$, $\theta_x = 0.3$, $r_x = 0.02$. **B** – effect of the background epistasis ($\theta$) on $\lambda$, for $k = 3$ (solid line) and $k = 7$ (dotted line), shown next to the load. $\theta_x = 0.9$, other parameters are as in A. **C** – when the background and the modifier epistasis parameters are simultaneously varied $(\theta_x = \theta)$, selection for the modifier becomes very sensitive to the recombination rate between the modifier and the introgressing block ($r_x$, values are shown corresponding to the lines from top to bottom). $S = 0.7$, $\theta = 0$, $S_x = 0.2$, $\theta_x = 0.7$, $r = 0.01$, $m = 0.003$, $k = 5$. **D** – selection for modifier increases almost linearly with the recombination rate between genes: each line represents $\lambda$ as a function of $r$ in the range of [0.001,0.0016] for the number of genes ($k$) varying from 2 to 7. $r_x = 0.02$, other parameters are as in C.